\documentclass[nofootinbib,preprintnumbers,amsmath,amssymb,twocolumn]{revtex4}

\begin{document}

\title{To the rescue of Copenhagen interpretation}
\author{Igor Salom}
\affiliation{Institute of Physics, Belgrade \\ University, Pregrevica 118, Zemun, Serbia }

\begin{abstract}
A recent paper "Single-world interpretations of quantum theory cannot be self-consistent" by D.\ Frauchiger and R.\ Renner has attracted a considerable interest of a broader physics audience and shortly elicited a number of replies. In spite of the objections that ensued, we find that significant part of the controversial initial claim has not been refuted - on the contrary, arguments presented both in the initial paper and in the replies seem to leave no longer room for the basic Copenhagen interpretation of quantum mechanics. We revisit the controversy trying to pinpoint the exact phenomenon that lies in its core, pointing out to some aspects of the problem that seem to have been mostly overlooked. Taking all conclusions into account, we here propose novel ways to approach the ostensible paradox, in which the Copenhagen interpretation is naturally preserved. Whereas arguments presented in Frauchiger and Renner paper and the replies that followed revolve over whether Wigner's friend would perceive only one or both of the possible outcomes, we offer the third, pretty obvious solution - namely, that Wigner's friend, in the well known setup, would subjectively perceive {\it none} of the potentially conflicting outcomes. As a side note, we also argue that, contrary to the often stated beliefs, hypothetical real experimental realization of Wigner's friend thought experiment actually would not provide us with any new insight into the quantum mechanics.
\end{abstract}

\maketitle

\section{Introduction}

After perusing the sequence of papers initiated by article \cite{MW} from D. Frauchiger and R. Renner, and followed up by papers \cite{OC1, OC2, DBB} of V. Baumann, S. Wolf, A. Sudbery and A. Hansen, a careful reader is left under the impression that a sort of silent mutiny has taken place in between of the traversed lines. In spite of the certain aftertaste that the main claims of the initial paper have been refuted in the replies, after the dust has settled it seems that the options left in the discussion actually no longer leave any room for the "standard" Copenhagen interpretation of quantum mechanics (QM). And notwithstanding the confronting presented arguments, our impression is that these replies have just given a more solid ground at least to a part of the controversial initial claim.

The paper \cite{MW} singles out one interpretation, i.e.\ many-worlds (MW), as the correct one, claiming to present arguments denying all the other available interpretations, even the "standard" Copenhagen interpretation (CI) of quantum mechanics. And where the subsequent replies successfully refute the main claim of uniqueness of the many-world interpretation, they actually make the case only worse for the CI. Namely, leaving aside (historical) questions what the Copenhagen interpretation exactly implies and no matter how vaguely it is defined, it is certainly agreed that CI postulates two things:

I) unitary evolution of the system wave-function in between of the measurements;

II) wave-function collapses upon the measurement.

In addition to this, it is beyond dispute that CI sees quantum mechanics as a complete theory, leaving no room for hidden variable interpretations or formalisms.

However, as we inspect more closely below, interpretations (or formalisms, in the terminology of \cite{OC2}) that adhere to both of these tenets seem to be effectively ruled out by both proponents and the opponents of the paper \cite{MW}. The attempt to refute CI as a possible interpretation of quantum mechanics is unconcealed in the paper \cite{MW}, with its aims clearly stated already in the title. The case for the CI is certainly not improved by the paper \cite{DBB}, which advocates a variant of the De Broglie–Bohm (DBB) interpretation (being a nonlocal hidden variable theory, this view is at obvious odds with the CI). However, the case is less clear, but not less so, with the papers \cite{OC1} and \cite{OC2}, which, at a first glance, tend to take no side of any particular quantum mechanical interpretation in the discussion, but try to merely refute or soften the strong claims of the Frauchiger and Renner paper.

Yet, they essentially conclude that this paper \cite{MW} has actually demonstrated the inconsistency of subjective collapse theories, leaving just two options: "objective collapse" and "no collapse" formalisms. Though these two might seem not only mutually exclusive, but also as exhausting all options, the type of "objective collapse" implied in this paper is such that it leaves no room for CI. Namely, in the objective collapse cases considered, subsystem consisting of Wigner's friend \cite{Wigner} and the measured particle does not evolve unitarily, in spite of being isolated. Such a collapse presupposes that the subsystem evolves according to some, possibly yet unknown, form of modified dynamics (non-hermitian Hamiltonian, explicit nonlinearities, etc), which spontaneously and irrespective of any measurement takes the state of the system into the new, collapsed state. Therefore, not only that evolution is in this case non-unitary, but there is also no true collapse in the sense of the additional postulate II - the collapse instead appears only "effectively", induced by the non-unitary evolution. This type of models, which predict new physics and obviously depart both from I and II (while postulate II can remain as an approximate rule) we will call {\it mechanical collapse} models (GRW and Penrose attempts are the most known models of this sort). We leave the notion of {\it objective collapse} models for a wider class formalisms where collapse still exists (there is neither practical nor ontological need to forever keep the terms in the wave-function that turned to be inconsistent with measurements) and yet are not necessarily of the mechanical collapse kind.

The other, "no collapse" option, removes the collapse postulate altogether. And no option seems to remain that could comply with the foundational definition of quantum mechanics, given by both I and II. A possibility that the subjective collapse (where Baumann and Wolf include also relational quantum mechanics \cite{Rovelli}, QBism \cite{QBism} and alike \cite{Brukner}) gives rise only to "apparent paradox" and yet not to a "scientific" one is indeed analyzed in \cite{OC2}. But the final verdict was that the "subjective collapse" in the basic Wigner's friend experiment with signaling seems to lead to a measurable contradiction.\footnote{Under a proviso that information between Wigner and his friend can be interchanged, which is in \cite{OC2} questioned only in passing.} Thus the silent consensus of the papers is that quantum mechanics, in its formulation from the first half of XX century, is, at least, incomplete theory (the latter if we allow for the mechanical collapse). And it is not only these papers: it is our impression that prevailing opinion of fellow physicists is becoming that either there is (objective) mechanical collapse, or the many world interpretation alternative.

However, this is not the first time that the incompleteness or incompetence of the standard form of QM was suggested. All such attempts to prove logical inconsistencies of QM in its basic form so far turned out to be futile, with the most notable and famous example of the EPR paradox paper. And often (like in the EPR case) such attempts did instead shed some light to previously less understood aspects of the quantum theory. We believe this could be also the case this time.

We next take a closer look at the arguments presented in these papers, pointing to certain overlooked aspects, after which we offer a third solution (or a third class of solutions) - to our knowledge not clearly mentioned by any of the established interpretations of QM. A solution that we believe is in accordance with the Copenhagen postulates and yet does not lead to contradictory predictions. It comes at the expense of further sacrificing certain intuitive concepts that we unwarrantedly used to take for granted (while we here maintain some other, disputed by alternative attempts to preserve CI), but this by now should not be a surprising feature of quantum mechanics in general.

\section{Recounting the arguments and counterarguments}

A the core of the Frauchiger and Renner paper is a cleverly devised setup that includes two "observed observers" (aka Wigner's friends). In spite that we certainly praise the ingenious play of probabilities and logical implications concocted in \cite{MW}, we agree with the position of \cite{OC1, OC2} that the essential paradoxical feature to which Frauchiger and Renner point out is already present in the basic Wigner's friend thought experiment. (The more complex setup of \cite{MW} merely demonstrates the same point in a more drastic way). Therefore, we will mainly focus on the original Wigner's friend experiment that was again brought into the spotlight by this series of papers, while only reflecting on the extended setup of \cite{MW} in the end.

The idea of "Wigner's friend" type of (thought) experiment is quite old, dating back at least to Wigner's 1961 paper \cite{Wigner}, though the same concept is found in the printed long version of Everett's thesis \cite{Everett2} which was, purportedly in the same form, originally written back in 1957. The well known thought experiment contains a sentient observer - i.e.\ the "Wigner's friend" - performing a measurement on a quantum system, while being ideally (informationally) isolated in a sort of a box. After the friend has done his\footnote{Since we refer to the original Wigner's argument, we will follow the suit and address the friend in masculine gender. Besides, the experiment itself does not seem to be very safe for the Wigner's friend, so we find it unsuitable to assign that role to a lady.} measurement, another observer - i.e.\ "Wigner" - has hypothetical ability to perform any quantum measurement on the joint system in the box (that comprises his friend and the initial quantum system that was measured in the first place).\footnote{This ability is sometimes called "the full quantum control", i.e.\ it is said that "Wigner has the full quantum control over the system in the box".} If the two measurements correspond to different bases of the initial quantum system, the thought experiment leads to intricate conclusions and superficial paradoxes.

Namely, let us, as usual, suppose that the initial system is a particle of spin 1/2, prepared so that its spin is aligned along x-axis:
\begin{equation}
|x+\rangle =\frac{1}{\sqrt 2} (|z+\rangle + |z-\rangle),
\end{equation}
where $|z+\rangle$, $|z-\rangle$ denote states of spin projection equal to 1/2, -1/2 along the z-axis. The Wigner's friend measures the z projection of the spin. If we assume that the wavepacket reduction (i.e.\ wavefunction collapse, measurement update rule) has taken place, the system in the box ends up either in a state $|F+\rangle|z+\rangle$, denoting the Wigner's friend who just saw the + outcome together with the particle with the spin along the z axis, or in the state $|F-\rangle|z-\rangle$, corresponding to the Wigner's friend who has measured the negative spin projection and the particle that is aligned accordingly. However, from the Wigner's perspective, all that is in the box is, at some level, just a extremely complex but isolated quantum system that, according the to rules of quantum mechanics, has to evolve unitarily, and the measurement act of his friend could result in no more than a mere entanglement of the state of his friend with the particle spin state. From his perspective thus the system must be in the superposition:
\begin{equation}
|W+\rangle = \frac{1}{\sqrt 2} (|F+\rangle|z+\rangle + |F-\rangle|z-\rangle),
\end{equation}
where $|F\pm \rangle$ is the state of the friend having observed $\pm$ (any apparatus that he had to use for the measurement is seen simply as an extension of the Wigner's friend himself, and its state is included in $|F\pm\rangle$). To make the difference objectively manifest, we assume that Wigner can perform the measurement on his friend and the particle in the basis $\{|W+\rangle = \frac{1}{\sqrt 2} (|F+\rangle|z+\rangle + |F-\rangle|z-\rangle)$,$ |W-\rangle = \frac{1}{\sqrt 2} (|F+\rangle|z+\rangle - |F-\rangle|z-\rangle)\}$. Then, if the friend is in either of the collapsed states, he will equally likely obtain $|W+\rangle$ and $|W-\rangle$ outcomes, whereas if the evolution inside the box was indeed unitary, then the chance of obtaining $|W-\rangle$ result is in principle exactly zero.

The essence of the paper \cite{MW} argument translates (oversimplifiedly, we admit) in this case to the observation that a way to reconcile these conflicting viewpoints (the only way, according to the authors) is to accept that both outcomes of the Wigner friend's experiment are equally (ontologically) real, i.e.\ that the friend must have both observed $|z+\rangle$ result and observed $|z-\rangle$ result, each in the corresponding "world" of the many-worlds interpretation.

The criticism of the \cite{OC1} and \cite{OC2} papers points out that the above argument inconsistently uses the "collapse postulate" of quantum mechanics while these authors make distinction between, according to them, altogether three possibilities:

1) The collapse is objective and objectively happens after the first measurement (or in its course) and thus Wigner has no reason to expect to see the interference result (i.e.\ to observe no $|W-\rangle$ occurrence). Indeed, experimentally Wigner would then measure both $|W+\rangle$ and $|W-\rangle$ outcomes with equal probability and there would be nothing paradoxical at all;

2) The collapse does not happen at all, which is the conclusion Frauchiger and Renner in \cite{MW} are seeking to derive, albeit this possibility is thus no longer the only one;

3) The collapse is "subjective", which corresponds to the above difference in measurement predictions of Wigner and his friend and thus, the authors agree, eventually leads to certain inconsistencies, so it is the possibility that can be, with some caution, refuted in this way.

As the consistent possibilities remain, according to \cite{OC1, OC2}, both 1 and 2, which are observationally different in principle (rendering these two different "formalisms" and not different "interpretations", as the authors rightfully insist).

The authors of \cite{DBB}, on the other hand, point to a generalized version of DBB interpretation as an explicit counter example for the \cite{MW} conclusions. In the terminology of the options 1,2 and 3, which are exhaustive according to the authors of \cite{OC1} and \cite{OC2}, this counter-example, in spite of not being the "many-world" interpretation, still fits into the second class of "no collapse" formalisms. (As some critics of the De Broglie–Bohm interpretations would say, DBB can be seen as many-world interpretation "with a pointer".) The flaw in the reasoning of paper  \cite{MW} that allows for the counter example presented in \cite{DBB} is the fact that Wigner's friend type of arguments can be used only as the motivation for giving up the collapse postulate. This leaves room not only for the "many worlds" interpretation, as the authors of \cite{MW} tend to conclude, but at least also for the more general original Everet's "relative state" interpretation and the Bell's pilot wave interpretation of \cite{DBB}.

But, in all this discussion, where is the Copenhagen Interpretation? V. Baumann, A. Hansen and S. Wolf are notably surprised in \cite{OC1} to find that the authors of the paper \cite{MW} have identified the "standard interpretation" of quantum mechanics with the "subjective collapse model", i.e.\ with the option 3, which, as all of them agree, does indeed lead to inconsistencies. In turn, they refrain from explicitly specifying which interpretation they consider as the "standard one", maybe simply in avoiding to assume any precedence or prevalence among interpretations. Nevertheless, we still take that by the "standard interpretation" all parties must consider what is more often called the Copenhagen interpretation (CI), as this is the only thing that can deserve such a title.

On the other hand, in the introductory section of \cite{OC2} authors announce they will treat "the standard quantum mechanics" as having different experimental predictions from the many-world approach in the case of encapsulated observers. In the context of their analysis, it then seems clear that the authors take the standard quantum mechanics to be one of the mechanical-collapse type, which we find unusual. In any case, CI obviously is not of the mechanical-collapse type, nor of the no-collapse type, and thus we are left under impression that CI is definitely disputed as a valid interpretation/formalism of quantum mechanics.

\section{The root of the paradox}

Let us take a closer look to where such a strong statement - as refuting of the CI certainly is - actually stems from. To this end, we will first consider a simplified version of the thought experiment, where Wigner's friend is replaced by a vastly simpler quantum system, e.g.\ with a single photon (we will not use another spin 1/2 particle to avoid confusion). Let a measuring device contained in the box produce a vertically aligned photon if the z-axis spin measurement outcome is +1/2, and let the polarization be horizontal otherwise. More precisely, and to be more in line with the photon mimicking Wigner's friend, let the photon be pre-existing, e.g.\ in vertically polarized state, with its state only switched to horizontal by the measuring device in the case of -1/2 outcome.

In this version, the Wigner's thought experiment is far less mysterious. Before the "measurement", the state of the "photon friend" together with the spin 1/2 particle is:
$$|\psi_0\rangle = |V\rangle|x+\rangle = \frac{1}{\sqrt 2} |V\rangle(|z+\rangle + |z-\rangle),$$
whereas after the measurement which, in this case without doubt, has merely introduced the entanglement, we have:
\begin{equation}
|\psi_1\rangle = \frac{1}{\sqrt 2} (|V\rangle|z+\rangle + |H\rangle|z-\rangle).
\end{equation}
Of course, we could above explicitly include the state of the apparatus/isolated environment, but the supposition is that these would simply factor out, and thus effectively change nothing of the analysis.

Once Wigner opens the box and measures the system in the W+/W- basis (here: $|W+\rangle = \frac{1}{\sqrt 2} (|V\rangle|z+\rangle + |H\rangle|z-\rangle)$, $|W-\rangle = \frac{1}{\sqrt 2} (|V\rangle|z+\rangle - |H\rangle|z-\rangle)$, there is no doubt that only $W+$ outcome would be possible. Such an experiment (or some variant of) is even practically feasible. Furthermore, there is no formalism or interpretation of quantum mechanics that would have trouble to explain such an outcome (otherwise, it could hardly qualify as a formalism/interpretation). Thus, there is no paradox on this level.

Let us also note the following: if Wigner, after opening the box and after his W+/W- measurement, decides to subsequently also measure the photon friend "memory" state (i.e.\ if he measures the polarization in the H/V basis) it is clear that he will obtain either vertical or horizontal polarization, with equal probabilities. The spin of the particle will, of course, become oriented in accordance with the photon polarization: up in the case of vertical polarization and down otherwise. However, this particular outcome tells us nothing about the particle spin alignment prior to this measurement of the photon in the H/V basis, and thus nothing about the spin orientation before W+/W- measurement and before opening of the box. This holds irrespectively of whether the H/V measurement was performed by Wigner on purpose, or it somehow effectively took place due to decoherence (supposing the environment for any reason prefers H/V basis). This conclusion, though obvious in the case of the photon friend, will be important to remember when we return to the human Wigner assistant.

Now, let us gradually increase the complexity of Wigner's friend and carefully note when the controversy occurs.

If we replace Wigner friend with couple of correlated photons/particles (e.g.\ with $|V\rangle|V\rangle|V\rangle|V\rangle$ encoding z+ outcome and $|H\rangle|H\rangle|H\rangle|H\rangle$ encoding z-), nothing is expected to change. However, as the number of particles playing the role of Wigner's friend grow, models that predict existence of mechanical-objective collapse will generally diverge in predictions from models that imply universal validity of quantum mechanics in its standard form (based on postulates I and II). At some point nonlinear/nonunitary dynamics that such models by definition have to predict must kick in and result in objective breakdown of the superposition of macroscopic states corresponding to the two outcomes. Whichever mechanism is responsible for such effective decoherence (i.e.\ mechanical-objective collapse) it is quite certain that its effect would have to become observable (via appearance of non-zero probability of W- outcome in W+/W- measurement) already on much lesser complexity scale than it is the complexity of a human Wigner's friend. Even in a far-fetched hypothesis that consciousness somehow objectively induces the collapse, it is hard to conceive such a nonunitary evolution that would require exactly human consciousness for the feat. On the contrary, it is more than plausible that the same physically-measurable effect could be then, at least partially, induced also by mind/brain of some animal, insect, by some key part of an (animal) brain organ, or even to some extent by a single neuron.

In other words, while the mechanical collapse models (or formalisms, as called in \cite{OC2}) would predict break-down of orthodox quantum predictions somewhere along the road from the single photon Wigner friend to the human Wigner friend, it certainly would not take actual experiment with humans to prove or disprove mechanical collapse theories. If much simpler experiments than these would not reveal hints of objective collapse, we could be pretty confident to abandon such models long before we are capable to perform literal Wigner's friend experiment. On the other hand, if mechanical collapse models are right, we would anyhow find it out by more feasible experiments much sooner than we become technically capable to perform the exact Wigner's friend experiment. Thus, having or not ability to actually perform Wigner's friend experiment does not influence our ability to tell apart mechanical collapse models from the rest. Likely, we will later argue that actualization of this experiment is not, contrary to often stated beliefs, necessary for gaining any new understanding of quantum mechanics.

In any case, objective collapse models imply existence of new physics, are obviously experimentally distinguishable from the standard quantum mechanics (postulates I and II) and are not the topic of this analysis. Our primary goal is to see if and how Wigner's friend experiment can invalidate Copenhagen formulation of quantum mechanics.

But, from the view of Copenhagen interpretation, adding new particles and increasing complexity of the Wigner's photon friend should not essentially change anything, thus the W- outcome must be as impossible with a human-like complexity of Wigner's friend apparatus as it was in the case of the photon friend. And, if there was no conceptual problem with a single photon friend it is a very intricate question how and at which point any paradox can arise as the this complexity grows. In particular, it is obviously essential to pinpoint which is that key element that introduces such a qualitative twist - from no problem at all, to a paradox as deep as to shake the foundations of quantum mechanics. It is hard to imagine how any analysis of this experiment can be carried out while glossing over this question - and yet this seems to be the case with basically all of the papers from this series.

According to \cite{OC2}, hint of this key element is nonchalantly given in parentheses at the end of the third section, where it is implied that Wigner's friend must be a "thinking (or computing) entity", as it must be capable to calculate and predict probability for Wigner to obtain result W-. While the word thinking evades easy technical definition and thus cannot really help us here, the word "computing" has much more clear meaning. It is basically defined\footnote{E.g. follow Wikipedia definition.} as an algorithm that performs a "calculation", which is in turn a "process that transforms one or more inputs into one or more results". However, what is the exactly the algorithm that should calculate probability of W-, and, more importantly, what are the inputs for such a calculation?

The only relevant {\it objective} input we have is whether the {\it objective} part of the measurement has taken place or not. And, unless something malfunctioned in the apparatus, the "objective part of the measurement" - i.e.\ the process of establishing a correlation between measurement outcome register and the particle spin being measured - will certainly have taken place. As we are not interested in malfunctioning apparatus, we realize that the input "whether the measurement has objectively taken place or not" is always true (or 1). Yet, the always constant input is as good as no input at all. Without variable input, i.e.\ with constant input, the only possible algorithm outcome can be also a beforehand given, preset constant, sealed by the implemented preprogrammed algorithm. Thus, we must implement from the start, as a constant, our prediction what the probability of W- outcome  will be, either 0 or 1/2. It is just as good as if we add another bit of information, e.g.\ another photon to our photon friend, but this additional one will simply be constant from start till the end of the experiment. Can adding of another, noncorelated photon in the plot bring about any paradox - certainly not! To conclude - increasing complexity of Wigner's friend to the level of a "computing entity" is not sufficient {\it per se} to introduce, even less demonstrate any paradox.

So, does the fact that universal quantum mechanic (contrasted to hypothetical mechanical collapse models) definitely predicts zero probability for W- outcome, and the fact that there is nothing objectively paradoxical with that means that there are no problems and potential paradoxes at all? Definitely not. However, we must recognize that the seemingly paradoxical situation arises only when we include subjective aspect in the picture.

Indeed, the problem arises only when and if we take into account that Wigner's friend subjectively perceives only one measurement outcome, either z+ or z-. Let us revise the original experiment taking account of both its "objective" and "subjective" aspects.

First, let us recount what objectively (by which we here actually mean subjectively from Wigner's perspective) must take place in this type of experiment. At some initial time $t_0$, Wigner's friend is, together with apparatus and the particle with spin along x axis, encapsulated in that isolated environment, i.e.\ the box. The corresponding state of the Wigner's friend with the particle is given with:
\begin{equation}
|\psi_0\rangle = \frac{1}{\sqrt 2} |F_0\rangle(|z+\rangle + |z-\rangle).
\end{equation}

At some moment $t_1$ the measurement, i.e.\ establishing of the correlation, takes place inside the box, bringing the state to the form:
\begin{equation}
|\psi_1\rangle = \frac{1}{\sqrt 2} (|F_1+\rangle|z+\rangle + |F_1-\rangle|z-\rangle).
\end{equation}

Here $|F_1\pm\rangle$ denotes the state of Wigner's friend at time $t_1$ that has just registered outcome $\pm$.
At some later moment $t_2$, the system is in the state:
\begin{equation}
|\psi_2\rangle = \frac{1}{\sqrt 2} (|F_2+\rangle|z+\rangle + |F_2-\rangle|z-\rangle),
\end{equation}
when Wigner carefully performs the delicate interference demonstration experiment, measuring the combined friend+particle state in the basis $W+/W-$ (where $|W+\rangle = |\psi_2\rangle = \frac{1}{\sqrt 2} (|F_2+\rangle|z+\rangle + |F_2-\rangle|z-\rangle)$ and $|W-\rangle = \frac{1}{\sqrt 2} (|F_2+\rangle|z+\rangle - |F_2-\rangle|z-\rangle)$  ). Here $|F_2\pm\rangle$ denotes the state of Wigner's friend evolved to the time $t_2$, with all corresponding memories and perceptions of someone who had measured spin projection $\pm \frac 12$ at time $t_1$. Necessarily, Wigner obtains W+ outcome with certainty, thus not altering the state of the combined system at all! Sooner or later after that, at some time $t_3$, Wigner's friend is finally allowed to leave the box, interact with global (i.e.\ Wigner's) environment and, either by telling Wigner about his experience or, more likely, by some decoherence mechanisms before that, he is effectively measured in the basis $F_3+/F_3-$. The outcome is either $F_3+$ and the overall new state:
\begin{equation}
|\psi_3\rangle = |F_3+\rangle|z+\rangle,
\end{equation}
which is the friend at $t_3$ knowing that he measured spin projection +1/2 at time $t_1$, accompanied with the spin vertically aligned, or $F_3-$ and the state:
\begin{equation}
|\psi_3\rangle = |F_3-\rangle|z-\rangle,
\end{equation}
that denotes him knowing that he measured projection -1/2 at $t_1$, while the particle has negative spin projection. Both outcomes come with equal probabilities of 1/2. This is exactly the same as the subsequent measurement of the photon friend polarization that we discussed earlier, which, in spite of the correlated state, does not really mean that particle spin had a well defined projection along z axis at time $t_1$. Therefore, no matter that Wigner's friend can clearly recollect himself measuring the spin and getting a very definite result at $t_1$, even being able to specify that result, this pose no contradiction whatsoever, since this result does not mean that z projections of the spin (and of Wigner's friend) was truly defined until the moment $t_3$.

We are surprised to note that all considered analyses of the Wigner's friend type measurement ignored to take into account the decoherence moment $t_3$, obviously finding it inessential. On the contrary, in our viewpoint this element will turn out to play the crucial role.

Anyhow, above we concluded that the standard quantum mechanics predicts that Wigner will certainly obtain W- result, while his friend at $t_3$ will nevertheless be able to attest that he obtained a definite result of the z spin component at the exact time $t_1$, and will be able to specify that result. As the matter in fact, we remind again that Wigner's interference measurement (in principle at least) did not alter the state of his friend at all (nor of the particle spin), so that the friend will have the same subjective observations and impressions as if there was no Wigner's meddling at all: just as if he simply measured the spin at $t_1$ and just left the insulating box at $t_3$. If the Wigner had ability to perform the interference measurement without his friends knowledge (e.g.\ during his sleep) the friend would have no way to find out if the measurement was performed or not. But, knowing about the interference measurement and of the outcome probabilities, from the friend's viewpoint there {\it seems to be} something wrong with quantum mechanics: he vividly remembers to have measured the z spin projection and obtained the definite result z+ (or it could have been z-) at time $t_1$, which means that both his state and the spin state should have collapsed accordingly. In turn, this implies that Wigner should have had 50-50 chances of getting either W+ or W-, but this is not what happens. And yet, this predicted confusion of the friend is fair and square in line with idea of universality of quantum mechanics, and there is nothing truly paradoxical in this account of the events. Interpretations respecting universality of unitary evolution (including CI, MW, DBB etc) thus predict that Wigner's friend must be puzzled if he considers events solely from his point of view, and that there is nothing really quantum-mechanically surprising, unexpected or even less contradictory about this.

To get any contradiction, we must ask what did the Wigner's friend subjectively experience at the moment $t_1$, or in between of $t_1$ and $t_2$. For, if he had subjectively perceived a single outcome at $t_1$, e.g.\ spin up, then he must have been entitled to apply the measurement update rule - i.e.\ to conclude that in the reality the spin was from then on oriented only positively along z axis - and thus to erroneously conclude that Wigner must have equal probability to measure W- as the probability to measure W+. The "single-spin-orientation" reality that Wigner's friend perceives at $t_1$ is then, at least in some sense, in contradiction with the "superposed-spin-orientation" reality that Wigner confirms in the interference experiment (why these two "realities" cannot coexist is in detail explained in \cite{Brukner}). And, on the other hand, we do know that he must have perceived a single outcome, since our subjective experience seems not be capable to be in superposition. Alternatively, if the friend is not allowed to use the collapse postulate (II), upon clearly perceiving a single measurement outcome - then it is very unclear who can claim that right, and we must conclude that the postulate is altogether pointless and that it should be discarded.

Actually, we can pinpoint the source of the paradox even further. On one side, we assume that Wigner's friend can be considered as a quantum-mechanical mechanism/device undergoing a unitary evolution. His brain is then essentially similar to a quantum computer (or anything more complex but nevertheless following the law of unitary evolution). Since we consider the friend jointly with his sealed environment, there is no true decoherence and the evolution is entirely unitary - there is no physical cause for any wave-function collapse (we have put the mechanical-collapse models aside). Next, we imply that such quantum-unitary brain-device gives rise to the subjective experience. Finally we attribute a novel capacity to this subjective experience: to effectively break down the superposition by "experiencing" only one of the superposed states that form the state of the parent brain-device (in some predefined basis). In other words, we expect it to induce the collapse of the wave function, in spite of the unitary nature of the mechanism that produces it. This does not seem mathematically reasonable and no wonder that this results in paradoxes. On the other hand, we cannot help to do so, and we do it with a good reason: in our subjective experience we always perceive well defined measurement outcomes, and concept of superposed outcomes seems utterly impossible. Alternatively, if we deny occurrence of the collapse even in this stage (and try to reinterpret subjective experience as do proponents of the many-worlds interpretation), then we must part with the postulate (II) and also with the Copenhagen interpretation.

Due to negligence of this point, essential in our view, in the most discussions on the subject, we feel an urge to emphasize it even more: to get any paradoxical feature it is absolutely essential to take into account {\it subjective experience} and not mere {\it intelligence} (i.e.\ reasoning, ability to perform computations, etc.).\footnote{It is not rare that these two concepts are confused, although it is not clear if and how much they have in common at all.} Nowadays, it is easy to imagine an artificially intelligent either classical or quantum computer that could mimic to some extent behaviour of the Wigner's friend. Based on the measured spin projection (and possibly some other input), this computer could produce the same output that the friend would produce, regardless of the output complexity (e.g.\ later account of the events, any elaborate messages issued before or after the interference measurement and alike). However, if resulting in a well defined output when provided a well defined input, then given a superposition of two spin outcomes the computer would necessarily end in the state which is a superposition of the corresponding outputs. (Since the environment is informationally isolated, this will also hold for the system of classical computer+enviroment, not only for a quantum computer). This is maybe more intuitive now than in the case with the Wigner's friend, though no different. Consequently, it maybe more intuitive that nothing {\it objectively} paradoxical can occur here. It is only if we assume existence of subjective experience of this computer (similar to that of a human) and question what it perceives as the result of the spin measurement, that we can run into weird or contradictory conclusions. Note that we neither claim that such a computer would posses subjective experience nor that it would not - we just stress that without it, there is no paradox at all.

Can we simply ignore existence of subjective experience, as it is not necessary to explain any objective behaviour? We doubt so. In spite that the mere existence of subjective experience is almost certainly not objectively provable (we direct reader to extensive literature on relation of Turing test with subjective experience, and the related long standing debates), it is nonetheless an experimental fact, which every one of us can immediately experimentally confirm (ignoring this would be equal to resorting to solipsism). Likely, it does not change anything if we claim that subjective experience is some sort of illusion - in that case the existence of the illusion is experimental fact that must be taken into account.

A number of solutions to this contradictory situation was offered in the literature. Maybe the most obvious one, that sparked the whole discussion in \cite{MW} by the claim that it is the only possible one, is the option to indeed discard the collapse postulate and claim that, in spite of the fact that Wigner friend can perceive only a single outcome, there are two copies of the friend in different branches of the many-world, each seeing a different spin orientation. Similarly, the DBB variant discussed in \cite{DBB} also abandons the collapse postulate. Alternatively, QBists take a sort of agnostic view on this matter, claiming that "to provide one agent the ability to conceptually pierce the other agent’s personal experience" by definition should be out of the scope of quantum mechanics (which is in turn understood as "gambling manual" for estimating outcome probabilities from personal point of view) \cite{QBism}. Instead of simply staying silent on the question of the friend's perceptions, but to us conceptually not far from it, one can give up the idea of single factual reality, i.e.\ of objectivity of the "facts of the world" - implying that facts can exist only relative to an observer (or reference frame) \cite{Brukner}. In this view, a statement, e.g.\ an observation about a measurement outcome of some agent, that is true for one observer is not necessarily such for the other. According to Brukner in \cite{Brukner}, this approach is a way to deal with the question of friend's measurement outcome within CI: "Copenhagenist (can) take the position that {\it there are no facts of the world per se, but only relative to observers}". In \cite{Brukner2} he further builds upon this view, turning it into a no-go theorem that effectively posits the denying of the "facts of the world" as the only way out for the Copenhagen interpretation.

For this reason, we will analyse a bit more closely the treatment of the Wigner's friend experiment exposed in the papers \cite{Brukner, Brukner2}. This approach allows that the realities of different observers can be even contradicting each other in certain sense. More precisely, it finds legit for an interpretation (and sees the CI as such) to negate the forth postulate of the no-go theorem in \cite{Brukner2}, the one which states: "One can jointly assign truth values to the propositions about observed outcomes ("facts") of different observers". In this particular case, it means that in spite of the fact that Wigner's friend obtained a well defined and single outcome of the spin measurement, his conclusion about the spin property is not necessarily true from the Wigner's information-reference frame. Though fascinating possibility as it is, this is an unusual and relatively high price we have to pay in the ontological and philosophical sense - to give up notions of the general scientific truth and facts, ideas which we thought are well defined and understood. Of course, this can be easily dismissed as an aesthetical complaint, which we agree it is.

However, Baumann and Wolf in \cite{OC2} argue that approach in \cite{Brukner} still possibly yields what they deem as inadmissible "scientific contradiction" in the case of the Wigner's friend experiment. Basically, they point out that not only the friend can calculate the probability of Wigner measuring W- outcome and obtain a different result from Wigner (which would be ok because realities of different informational-reference systems cannot be compared), but also he can communicate these conclusions to the Wigner - which, if we correctly interpret, gives rise to what they call "scientific contradiction". In our opinion, the incompatibility of Wigner's and friend's predictions is not necessarily detrimental per se. As the matter in fact, we have already pointed out that all unitarity-respecting formalisms of quantum mechanics predict that Wigner's friend would have to conclude the probabilities of the interference measurement outcomes to be 50-50 percent, as long as he is judging solely from his own perspective. What this "scientific contradiction" makes maybe more of an obstacle in this case, is the insistence of Brukner in \cite{Brukner} on the complete symmetry between the observers - something we do not find fully appropriate and necessary in the particular given setup (since, as we have seen, only one of them here will be mistaken if considers the collapse and the events solely from his perspective).

Finally and most importantly from our viewpoint, it is not clear (or at least not clarified in \cite{Brukner}) what is the relation of the spin projection result the friend obtains (from his reference system) at $t_1$, and the potentially new outcome that the friend later remembers, decided at $t_3$? If these results coincide, then both Wigner and his friend later know what was the z spin projection at $t_1$, in spite of the interference demonstrated at $t_3$. Brukner, on the other hand, goes in length in \cite{Brukner} to explain that this cannot be so. Besides, why would this be different from the case of the "photon friend" where we concluded that subsequential measurement in H/V basis would not reveal electron spin projection as it was at $t_1$, but only as it became at $t_3$ (the opposite would radically contradict our usual understanding of quantum mechanics, and run into various inconsistencies). Even more strangely, if the final result stored in the friend's memory is not correlated with the one obtained at $t_1$, and could be in principle a different one, then why and at which point this flip occurs? We stress again that Wigner's interference measurement did not alter the friend+spin system in any way, so it is hard to see this as the cause of the flip. Not implying that these points disqualify the above interpretation as a viable explanation of the Wigner's thought experiment, some readers may find it to be a shaky argument for the CI case and difficult to take the solace from the author's claim that this is the only way to reconcile CI with Wigner's gedanken experiment. In this regard, we find that possible additional interpretations supporting CI would be quite welcome.

Indeed, now that we have pinpointed the root of the Wigner's friend paradox to be inseparable from the phenomenon of subjective experience, which is in turn a concept even less understood than the foundations of quantum mechanics, we have opened a door to new understanding of implications of the Wigner's friend experiment.

\section{Subjective collapse is subjective}

We believe that a resolution much more in line with the spirit of quantum mechanics, which follows the basic CI, has been overlooked due to implicit unnecessary assumptions made in the reasoning from the previous section. Namely, it is the possibility that, from the Wigner's perspective, the problematic friend's {\it subjective perception} of spin measurement at $t_1$ simply did not exist prior to $t_3$!

Actually, all we need in order to get there is to steadfastly follow the CI. All basic CI standard concepts and postulates can remain as they are: there is a single reality and there are universal facts of the world. The evolution is unitary, apart from the moments of collapse that coincide with those of measurement (measurement in a broader sense). In the terminology of \cite{OC1, OC2}, the measurement postulate is applied according to what the authors call "subjective collapse interpretation". However, it must be done consistently, which, in our opinion, none of the papers \cite{MW, OC1, OC2} does. Namely, all that Wigner's friend needs to do to get his predictions right is to refrain from attributing his "subjective" collapse to other observers (i.e.\ to Wigner). If he wants to infer the outcome of Wigner's experiment, he may, but he must look from the Wigner's perspective - and from that perspective there is no any collapse and the evolution of the system in the box is unitary. In that case, he would correctly calculate that the probability of W- interference outcome is zero. The friend is entitled to consider from his own perspective - which includes collapse - only subjective outcomes of his own measurements. He can predict that at $t_1$ he is going to measure the particle spin projection, and that he will observe a single outcome of two possibilities, with probabilities 1/2. And indeed, as we have shown, that is exactly what happens {\it from his perspective} according to all interpretations/formalism of quantum mechanics which respect the universal unitarity. If he follows the CI rules consistently in this way, both him and Wigner will get all the predictions right, and there will be no "scientific contradiction" at all. It only makes sense to call such a consistent application of subjective collapse the "subjective collapse" interpretation.

In this account of the events, we have basically literally followed the standard postulates of quantum mechanics. The only assumptions we must adjust in this approach are actually not related to physics, but to our understanding of the phenomenon of subjective experience (i.e.\ of consciousness). The part that runs against our usual intuitive assumptions is answering, from the Wigner's point of view, the question "what does the friend subjectively perceive in between $t_1$ and $t_2$?"

This question can be answered in multiple ways, each defining a particular variant, or a "sub-interpretation", of the "subjective collapse" interpretation (while the latter we see only as a natural reading of CI). The first, and a quite obvious one that we have already announced, is to accept that simply there was not any such friend's subjective perception in the period from $t_1$ to $t_2$ from the Wigner's aspect. Indeed, there is no any trace in the universe that such subjective perception ever existed. As we have already discussed, in the end the friend does have clear recollection of the events during and after the measurement, but the content of these memories was determined at the moment $t_3$. Even if subjective perceptions of a single outcome could be ascribed to the Wigner's friend at the time $t_1$, the corresponding memories would be erased by the interference measurement at $t_2$ and the memories - but possibly of a different outcome - would again appear after the decoherence at $t_3$. So, whatever Wigner's friend tells us after the measurement, it has nothing to do with his subjective perception in between $t_1$ and $t_2$. There is no any proof that his perception existed in this period (later we will address the potential problem of signaling). And, we have learnt in quantum mechanics that most often one should not assign definite reality to events/things of which exists no account/information. In addition, it is exactly this subjective perception that seems to cause the conflict with the standard postulates of quantum mechanics. Why is it then so widely implied that Wigner's friend must posses subjective perception from Wigner's viewpoint even in between of $t_1$ and $t_2$?

We believe that the main reason why the other option has not been analysed at all (to our knowledge) is a combined  effect of two assumptions that are tacitly taken for granted in this context. The first assumption is that the subjective perception is a mere product of brain functioning. In truth, while the correlation of the brain functions and the subjective perception (consciousness) is beyond any doubt, whether the subjective experience can be explained solely as the emergent phenomenon of the biological computations taking place in the brain - is much of an open question in cognitive science and philosophy of consciousness. Not only that the Wigner's friend discussion in the literature tend to, without saying, imply positive answer to this long standing question, but authors go even a step further: seemingly inadvertently, without saying or commenting, they imply that subjective experience must also emerge from and directly accompany even a "potential" state of a brain - i.e.\ a brain state that exists in a superposition. But, the latter assumption is so strong and far from being obviously warranted that even accepted DBB interpretation of quantum mechanics already does not adhere to it: unlike the many-worlds interpretation which assumes that any branch that contains a state of some operational brain also lends existence to subjective experience of the brain owner, for DBB interpretation wavefunction of the brain is not enough to yield subjective perception - only the "branch" filled with "actual particles" somehow produces subjective experience.

Thus, one way to understand the absence of friend's subjective experience (from Wigner's perspective) in the period from $t_1$ to $t_2$, is to accept that subjective experience pertains only to actualized reality - i.e.\ corresponds to collapsed states (and in this sense to classical information). "Undecided" or potential reality in such view does not give rise to subjective perception. That is, from Wigner's perspective, his friend's subjective perception becomes defined and thus real and existent only at the moment $t_3$, i.e.\ at the moment when his brain state also gets defined (collapsed). And when it gets defined at $t_3$, it contains all the memories and experiences pertaining to period from $t_1$ to $t_2$ that his friend will feel (and tell about) as if they happened not at $t_3$, but from $t_1$ to $t_2$ (we stress once more that this conclusion about friend's recount of the events is shared by all unitarity obeying formalisms). Nevertheless, before $t_3$, while the friend's brain is from Wigner's perspective in superposition regarding that part of experience from $t_1$ to $t_2$, there is simply no corresponding experience yet. In this very sense, and in contrast to the Wigner's "consciousness causes collapse", this view can be dubbed as "collapse causes consciousness".

Another possible way to understand why and how there can be no Wigner's friend experience in between $t_1$ and $t_2$ from Wigner's aspect, is to accept that from this perspective his friend just does not have subjective experience at all. Not only from $t_1$ to $t_2$, but never. Such a view could be dubbed as "no objective consciousness" interpretation. In other words, the subjectiveness of subjective experience should be taken to such an extent that not only it makes no sense for Wigner to speak or discuss his friend's subjective perceptions, but that he somehow must not assume their existence at all. In this view, subjective experience can exist only as subjective, and all the problems arise from in-principle wrong attempts to objectivise subjective perception. To grasp the full meaning and implications of this statement - if possible - lies beyond the scope of this paper. However, we must note that such a (somewhat radical) conclusion would also provide a curious explanation of (or at least an unexpected connection to) the long standing problem in the philosophy of consciousness, known as the "problem of other minds" (related to what seems to be inability in principle to infer whether or not a given entity/agent possesses subjective experience). From the observational point there are certainly no paradoxes nor conflicts at all - the situation is symmetric from the viewpoint of any of the observers, and each of them can correctly calculate outcome probabilities of any measurement (of course, each time it is necessary to take the corresponding viewpoint). In addition, lack of subjective experience of his friend from the Wigner's perspective certainly does not prevent his friend to act sensible in every aspect - in the very root of the "problem of other minds" is the fact (or at least the hardly disprovable assumption) that subjective experience is not a necessary prerequisite for anyone's objective behaviour.

There is another approach, at least superficially different, to answer the question of "what does the friend subjectively perceive in between $t_1$ and $t_2$?". This one requires to tweak a bit further our understanding of time, and of its relativity. Instead of the variant in which Wigner denies existence of his friend's subjective experiences in between $t_1$ and $t_2$, in this view he is actually able to assign well defined perceptions to the friend in this period - but the perceptions related to the outcome that is defined at $t_3$! That is, Wigner may conclude that during the period from $t_1$ to $t_2$ his friend, at each corresponding moment, is currently living through experience in which he has just seen a well defined spin measurement outcome and is pondering, at any given moment, its value in his mind. But this outcome that he perceived is not the one that corresponds to actual spin alignment at $t_1$, it is instead the result that {\it will be} defined only at $t_3$. In other words, in this view he sees the result that will be determined in the future! The advantage of this "perception of future" perspective is that the Wigner's account of temporal evolution of his friend's subjective experience can match what his friend later recounts of the subjectively experienced past (i.e.\ these two viewpoints become synchronized).

Nevertheless, shouldn't this be inconsistent, even ridiculous? Surprisingly, it does not seem so. It could potentially lead to paradoxes if the friend would be able to communicate the result he perceived "from the future". However, at the very moment he manages to send this information outside of his sealed environment, his superposition from the Wigner's point of view will collapse and the information becomes that of the spin at $t_1$.\footnote{We dismiss possible objection that Wigner could artificially engineer friend's future memories to bring contradiction with his "prophetic" perceptions, as requiring too many assumptions about the connection of the brain and subjective perception. But, if pressed to comment on this possibility, we point that this very inability of the friend to communicate the result, enables him to perceive - without contradiction - whatever future will become compatible with Wigner's meddling.}

Moreover, we feel that our persistence to speak about the precise moment at which this outcome is becoming defined is ill-suited to the situation.\footnote{Vaguely, this resembles the situation when Alice and Bob are space-like separated, share an equally polarized pair of entangled photons and both measure the polarization along the same axis: depending on the relative motion of the frame of reference, either Alice outcome will be random and Bob's predefined, or the other way around. The comparison here is only insofar as to remind that quantum mechanics often leads to counterintuitive situations regarding time and certanity/chance.} Positioning and ordering of the events in time might be a problematic concept itself if there is no flow of information that could provide synchronization. Particularly, ordering of the subjective perceptions might only have sense if these are correlated. Therefore, we are of the opinion that the "collapse causes consciousness" interpretation (one where the friend's subjective experience does not exist for Wigner in between $t_1$ and $t_3$ and then emerges at $t_3$, while he himself subjectively continually perceives living through this period) and the "perception of future" interpretation (in which the friend perceives the outcome "from the future" and Wigner is able to attribute continuous perception to his friend in between $t_1$ and $t_2$) - are not essentially different. There seems to be a mathematical equivalence of the two interpretations and, from our understanding, it points to a deeper relation between time and subjective experience - a connection beyond our present understanding.

Finally, let us address the important issue of objectivity of the "facts of the world". The acknowledged subjectivity of the collapse in (some of) presented views should not be confused with impossibility to assign universal Boolean values about the statements on (subjective) observations of different agents. With a (partial) exception of the "no objective consciousness" sub-interpretation, the other two options are more about temporal relativity of the subjective experience. Note that in none of the proposed (sub)interpretations are there any contradicting perceptions of the observers: conflicting outcomes that cannot be allowed to coexist indeed do not coexist. Thus it seems logically safe to take the subjective perceptions of agents as "the facts of the world" which are independent of the observers.\footnote{Of course, we here consider only sane, rational subjective perceptions.}

Next we will discuss some of the more delicate aspects of our explanation(s) and some of the potential objections to it.

\section{"When" and "who" of the collapse}
\label{sec:whenandwho}

The mystery of "when does the collapse occur?" is likely the most profound and elusive of questions in quantum mechanics, at least of those that can be put into a simple sentence. Answering this question relies also on answering of "what causes the collapse". Physics has matured past the expectations that trivial (at least philosophically) explanations can be given to these conundrums, so we cannot be expected to pull off some miracle either.

Yet, what seems to be maybe the most fascinating and puzzling aspect of this question, is the apparent symmetry and arbitrariness of the answer, that mobility of the position of Heisenberg's cut in the von Neuman's chain. It is this curious property that allows for so many vastly different interpretations of quantum mechanics, including the variants of subjective collapse approaches. A general agreement upon this exists at least as long as we do not consider "encapsulated observers" cases (of the Wigner's friend type). However, the authors of \cite{OC1, OC2} explicitly insisted that "subjective collapse" interpretations break down when faced with "encapsulated observers" type of thought experiments. Contrary to that, we have shown above a number of such (sub)interpretations that pose no contradiction even in this type of gedanken experiments.

Ostensibly it looks that the approaches we proposed rely heavily on existence of "subject" in the course of analysis, and that the trouble of defining eligible "subject" can lead to complications. Related is the problem of defining "whose" an experiment is, as we insisted that one should calculate the predictions exclusively assuming the appropriate point of view. Nevertheless, none of these is a true problem exactly due to impossibility to objectively assign who or what possesses the subjective experience, and who or what does not. There is an arbitrariness of decision to watch anything from the outside and see it as a quantum-mechanical clockwork device, or to assume existence of subjective experience and to deduce these inner perception outcomes.

For example, in the contemplated Wigner's experiment, we have already discussed possibility that the friend is not a sentient being, even replacing him with a mere photon. Regardlessly, Wigner obtains the same predictions about his interference experiment. He may, in addition, attribute quality of the subjective perception to his friend (was he a human, an animal, a computer, or even the photon) and if he does so, he can accurately predict what this subjective perception in the end will be - to the extent that his prediction can be corroborated by his friend (if the latter can give his subjective account of the events). Yet, never he can objectively prove whether his assumption to grant his friend the qualia of subjectivity made sense or not. And this inability certainly does not influence any objective prediction.

It is more intricate to consider a replacement for the Wigner. We can replace him with some elaborate device, and assume that the machinery was devised and put into action by Wigner's friend himself, prior to his isolation in the box. Since it is, to the best of our ability to name the things, now the friend the "one that performs the experiment", should not the fact that he subjectively perceives the spin outcome alter the conclusions and change the probability of W- interference outcome? Not in the least. Subjectiveness of the interference measurement reflects only in the moment when the friend decides to learn the outcome of the interference measurement. If both of the outcomes W+ and W- were possible (i.e.\ with nonzero probabilities), the friend would subjectively collapse this superposition. Up to that moment, the delicate "Wigner device" merely gets entangled with the system in the box, and with the quantum machinery of friend's brain, and gets into a superposition prone to later collapse by the friend's measurement (the fact that probability of one of the outcomes here is zero is inessential). Let us analyse in some more detail how the Wigner's friend would, at time $t_0$, see the events to come: "At $t_1$ my spin measurement apparatus within the box and my brain will get correlated with the particle spin. At $t_2$, a device named Wigner that I've prepared earlier and that stays outside of the box will carry out a routine called 'interference measurement'. Its two-state output register, with possible values W+ and W-, will get correlated with the entire system in the box, including my brain. {\it A priori}, there are four outcome possibilities: I perceive spin up and then at $t_3$ find the register in the W+ state, I perceive spin up and the register in the W- state, and the two corresponding possibilities with the spin down. When I use quantum mechanics to calculate probabilities of these particular outcomes, I get that the probability of W- in either case is zero, while the other two probabilities split into equal parts." (Note that the precise moment $t_3$ is not essential here.)

One could object that if the friend repeats his analysis and calculation in what he subjectively perceives as time between $t_1$ and $t_2$, he might conclude "I've just seen a well defined single outcome of (e.g.) spin up and thus the Wigner device measurement that will take place at $t_2$ should result in W+ or W- with equal probabilities". But this is again the same false paradox and necessary illusion of the Wigner's friend that, as we pointed out earlier, any unitarity respecting interpretation (including MW and DBB) must entail. The mistake is the result of failure to properly account for all relevant elements of the entire system.

What if we replace both Wigner and his friend with some machinery? Would there ever be a collapse? Does it make sense, and is there any necessity to attribute consciousness to this unspecified machinery? Now that is pure philosophy and metaphysics. At any rate, to get any objective readout and to discuss anything, we must involve a subjective experiencer at some point.

Indeed, it is weird but true that a "single" subjective experiencer (whatever that would mean) is, mathematically at least, enough to account for any objective measurement outcome. Thus a common objection is that "subjective collapse" approaches may lead to a solipsism of a sort.\footnote{As a curiosity, we mention an anecdote told by Coleman, that pondering on the measurement problem inevitably led him even to this option. He remarked to Aharonov his conclusion that quantum mechanics indicates that it must be only him in the Universe causing the collapse, and the latter replied with "Tell me, before you were born, could your father reduce the wave packets?"} Nevertheless, this is absolutely unnecessary metaphysical direction, and thus a pointless objection.\footnote{Instead, one could argue that this profound symmetry and arbitrariness of the cut position rather points to the underlying oneness of the subjective experiences, but we do not intend to endeavor into metaphysical discussions.}

Finally, we must also mention, at least as a logical possibility, that in spite of the existing mathematical symmetry between different observers and points of view, nevertheless there could exist a preferred "frame of reference", one in which "facts of the world" arise and are becoming determined in a time-orderly way. While this idea probably hurts our aesthetical feeling and does not bode well with the Occam's razor argument, it does have certain advantages. No longer there would be any trouble of how to reconcile "when" and "who" of the collapse from different perspectives. This informational frame of reference would then define the true reality, and the objective collapse. Our subjective perceptions would then emerge when the state of our brain becomes defined (actualized, or collapsed) within this frame. After all, situations of the Wigner's friend type are maybe too artificial, or are at least extremely rare, while apart from such cases the universe generally belongs to a single information-reference frame: as far as we know there are no truly informationally isolated parts in the Universe, and it might be that such a concept is for some reason unfeasible even in principle.

Nonetheless, we personally think that such a symmetry reduction is not truly necessary and that instead we should accept that there might be simply limits to our intuitive understanding (itself too much bounded by time-ordered thinking in terms of cause-and-effect) which make difficult our attempts to comprehend this perplexing symmetry.

To come to terms with this strange subjectivity and symmetry of the collapse, and yet objectivity of the "facts of the world" proposed in these views, the following (albeit imperfect) analogy might help. We might imagine a movie plot and that we are following the action. There is a man, Wigner, in his laboratory, and there is a story that goes on about the events there. Then a character that was not there before appears in the scene - Wigner's friend. What was not told before about the friend, we now find out and it is actually only now that this part of the story becomes defined and "real" - but only those details that are shown in the movie, while the writer of the script still has the freedom about the rest of it. The movie is created and the plot is written as the camera rolls.

Or we might film the entire scene letting the camera follow the Wigner's friend from the onset. Then he leaves the box and the Wigner enters the scene, and everything that was not fixed about the Wigner and his lab by the earlier part of the movie now becomes defined, and that part of the script written. The camera position and movie directing are different, but the script (facts of the world) in the end is the same, unique. To what extent and whether the script is fixed and written in advance, or to what extent (more likely) we take part in writing of the script as the story goes on - is rather a metaphysical question out of the present scope. The order in which the parts of the script are written does not need to coincide with the time as experienced by any of the characters in the movie, it might not even necessarily be commensurable with the internal time in the movie (e.g.\ if the friend decides to send an elaborate message to Wigner in between $t_1$ and $t_2$, as envisaged by Deutsch and discussed in details below, then this message-content part of the script is for Wigner written and experienced before the part about the spin orientation at $t_3$, while the friend experiences these two simultaneously).\footnote{Interestingly, the fact that the stories from different angles are consistent, might be indicative of that the plot is being written somehow jointly, or in some sense "at one place".} Personal awareness exists either only when the camera is in our hands ("no objective consciousness"), or it can be taken to arise subjectively for us also when we enter the scene ("collapse causes consciousness"), or the characters do have subjective experiences even when they are not shown in the scene, but are being told, if necessary, some future lines from the script ("perception of future").

The view in which there is actually no true symmetry of the observers (i.e.\ of the movie characters in the analogy) is conceptually far simpler to put in the movie context. In that case there is only one, well defined camera position (capturing most of the Universe), and the script is being written as the plot evolves. The movie shows the Wigner's laboratory and the Wigner himself - which Wigner "feels" as being aware. On the contrary, what happens with the Wigner's friend is not decided yet in the script, and there is {\it objectively} no subjective experience of the friend at that time. At the moment the friend jumps out of the box his part of the script has to be written, with all the history of the events in the box, which he then  subjectively experiences as living through these events.

Of course, we emphasise that the movie analogy is neither complete nor bounding (in the sense that the proposed interpretation are sufficient without and independent of philosophical connotations that the movie parallels may bring).

\section{Troubles with signaling}

We offered the alternative explanation of the Wigner's thought experiment assuming that there is no informational "leaking" from the friend's isolated system (box). But what if Wigner's friend takes care that some trace of his subjective experiences in the period after $t_1$ is left in the universe, i.e.\ by sending some deliberate meaningful information?

Idea of signaling in a Wigner-like thought experiment dates back at least to Deutsch proposal in \cite{Deutsch}. He envisioned that the person playing the role of Wigner's friend sends information on whether or not he obtained a single outcome of the spin measurement. The idea was that he is not sending any information about which of the two possible outcomes of his experiment he obtained, but merely the information whether he perceived a single measurement outcome or maybe some sort of superposition of the outcomes.

As in \cite{Brukner}, we believe that there is no any serious doubt that Wigner's friend can see anything but a clear single outcome - as opposed to any sort of "superposed measurement result". Despite the possible differences in understanding when the subjective experience occurs, i.e.\ about when the content of these experiences is determined, we do not expect these "later decided" experiences to be qualitatively different from those "normally occurring" (i.e.\ in conditions of the normal flow of information). More importantly, what Wigner's friend experiences when isolated in the box cannot depend on whether he is subsequently measured in the interference measurement by Wigner or not (this is especially so in the Deutsch and other interpretations in which experience arises and is determined before the interference measurement). If he is to perceive any superimposed or otherwise unusual measurement result, than a similar effect must befall any isolated observer upon any measurement of a non-eigen state. As such measurements and observations are rather a rule than an exception, it is likely that any isolated observer would in that case see some sort of blurred reality. If this idea is taken seriously, it is quite unclear where such fuzziness would end - when the information about the result reaches another human being (but why then Wigner's friend himself was not sufficient?), a number of humans, or what?

Besides, isolation does not have to necessarily be in the form of some closed surrounding as in an "information proof" box (which is hard to be practically realised) - such an effect should in principle be also experienced by anyone alone enough in the space: it would take time equal to L/c for any information to reach the closest (relevant) object (where L is distance to that object), and for that time some of the observations would be "blurred". This is obviously not recognized as a realistic peril of the space travel, and we believe with a good reason (depending on some further details, it is arguable that such effect would have been already noticed during Appolo missions).

Therefore: i) we certainly do not need to perform experiment as elaborate as the Wigner's friend one to rule out possibility that the friend can see superimposed measurement results - furthermore, we argue that it can already be ruled out; ii) possible message that Wigner's friend would send about seeing or not single reality thus does not truly carry any information (i.e.\ it is an always-the-same constant bit). From this viewpoint, sending this bit of information by the Wigner's friend is as good as sending a uncorrelated predefined bit which cannot change anything in the analysis.

And yet, not entirely. Let us say that Wigner's friend keeps to himself to deliberately choose the content of this bit of data (instead of automatically having it set to the constant which denotes "I saw a well defined single outcome"). Let's also suppose that Wigner then performs the interference experiment and obtains W- result, even if his quantum control of Wigner's friend system is ideal. Would such a result now signal a breakdown of the unitarity of quantum mechanics? No longer. It would indicate the nonzero probability that friend nevertheless {\it decides} to send the bit value correlated (or slightly correlated) with the spin orientation outcome that he measured. Maybe by a pure whim, he in the last minute decides to use this bit of information to uncover the obtained result. This is not same as a measurement error - this effect must be there as the consequence of the free will (or of what we perceive a such). If it was an automate or a deterministic computer instead od Winger's friend, this could not happen, but could the computer inform us if it {\it subjectively} perceived a single measurement outcome? It could maybe be argued that the probability of the friend not acting by the agreed rules could be calculated (a posteriori) by Wigner who knows all details about his friend's brain and its functioning. Nevertheless, it all introduces necessity for deeper and deeper assumptions about relation of the brain and subjective perception, of free will and computability. Are the free will and existence of subjective experience compatible with the determinism and computability? Should we go as deep as to discuss possible relevance of the Goedel incompleteness theorem? We feel that a physics discussion should try to stay away from these matters, as long as it is possible.

Thus far we believe to have shown that introduction of this communicated bit between the friend and the Wigner would not provide us any new insight about whether or not superimposed measurement outcome can be observed, but instead would introduce unwanted non zero possibility of W- outcome, while relying on many ad hoc assumptions about brain and consciousness.

However, reading Deutsch paper \cite{Deutsch}, one is left by impression that question of whether the friend sees superimposed reality is more of a rhetorical nature. It is rather there to stress that the friend is obviously aware of a single and only single outcome, and thus eligible to apply measurement update rule (i.e.\ to conclude that the collapse happened). Deutsch conceives that the friend, instead of sending a single bit, even elaborates on the topic and writes: "I, Professor X, F.R.S., hereby certify that at time t''' I have determined whether the value of the North component of the spin of atom is +l/2 or -1/2. At this moment I am contemplating in my own mind one, and only one of those two values. In order to facilitate the second part of this experiment, I shall not reveal which one." Such a complex message (written deliberately by a human) really seems to leave no doubt that consciousness, subjective experience and free will (whatever is exact meaning of these notions and if they can be objectively attributed to  anyone at all) must have been present and at work during the creative task of this writing. But, can it be so certainly assumed that the particular obtained result (+1/2 or -1/2) cannot influence at all contents of such a complicated message? At least by a single detail, even subconsciously - and yet enough to destroy the interference? If it was a deterministic computer instead of a human such a risk would not exist, but we are analysing the effects where a computer seems not to be sufficient. Can we be sure that this human-vs-computer aspect of the problem is only coincidental and not indicating some limitation in principle? Again, proceeding further with this level of communication between the encapsulated observers, forces us to imply certain ad hoc assumptions on the consciousness and free will even to a stronger extent than before. All cautions we considered when discussing a single bit message are only much more pronounced in this case.

Note that, to the contrary, as long as the friend's system is absolutely isolated (there is no any information flow outwards) and Wigner has the complete quantum control over it, there is no limitation in principle for Wigner to demonstrate complete impossibility of W- interference result.\footnote{For this it is also necessary that Wigner's friend has no possibility even to halt the experiment - otherwise, there is always some possibility that he could decide to do that depending on the outcome, which would in turn reflect on the probability of W-.}

Nonetheless, let us push forward and assume that Wigner's friend experiment can be carried out also in presence of this type of communication between the friend and Wigner (only at the cost of lesser "interference visibility"). What would be implications on our "collapse causes consciousness" view?

The premise is then that the friend manages to send to Wigner a meaningful message in between $t_1$ and $t_2$ which is {\it absolutely uncorrelated} with the particular measurement result. Then this aspect of the friend's consciousness is also absolutely unrelated ("untangled") from the obtained result and it indeed becomes real, determined and subjectively experienced between $t_1$ and $t_2$. The aspect of the subjective experience that pertains to the explicit spin projection outcome still remains to be "decided" - and thus to become a part of subjective experience - only at $t_3$. However, this does not mean that Wigner's friend has any unusual experience of contemplating about the message while having strange gaps in his thoughts/perceptions. According to the premise, it was only the particular spin outcome, and not the fact or perception of its unique existence, that stayed uncorrelated with the sent message - we see no reason why the latter could not be well defined without the former. On the other hand, at $t_3$ all the aspects of his subjective experience about the events from $t_1$ to $t_2$ are finally determined/decided and his perception and recount of the events has no sort of gaps. From his viewpoint everything would seem absolutely ordinary - he perceived an outcome and wrote a message. Counterintuitive as it may seem based on our usual assumptions on the consciousness and subjective experience, we do not see any actual contradiction in such an outline of the events.

Here a comparison to an artificially intelligent (quantum) computer might even help: it is devisable that such a machine could compose a message of the sort contemplated by Deutsch, in which the computer expresses the fact that it has performed the measurement and got a single result stored in a single bit memory register. It could report in the message, in earnest, that it is "contemplating within its registers one, and only one of those two values" - as truthfully as it is possible. For, whichever algorithm that can issue such a statement when only 0 or only 1 is written in this "measurement outcome" register, will detect no difference and will result in the same output statement also when the register contains a superposition of 0 and 1 - since, by the premise, the algorithm output is not correlated with the value of the register. Finally, receiving (i.e.\ measuring) such a message (by Wigner) would collapse a certain number of registers - all those involved in composing the message (which might be in superposition if the algorithm is not entirely deterministic). However, since the register containing the result was uncorrelated with the output message, its content would stay in superposition.

One could object that now we are replacing the friend with a computer, whereas we have already noted that validity of such replacement is tricky and dubious at best. Yet, we did it for a number of reasons: i) to underline once more that without considering friends subjective experience there is nothing strange, let alone paradoxical in this thought experiment, ii) that even introducing issue of subjective experience does not bring in any paradox per se - it is also necessary to imply certain assumptions about this subjective experience, in this particular case, to imply that it must behave differently than a (quantum) computer (in the sense to imply that subjective experience cannot be "collapsed" partially, unlike the computer registers), and iii) to help us by this analogy to accept possibility of otherwise quite counterintuitive hypothesis.

Therefore, the signaling argument cannot invalidate "collapse causes consciousness" view. The significance of the message sent by the Wigner's friend is even lesser for "no objective consciousness" interpretation. In this view no subjective experience can be attributed to the friend from the Wigner's aspect. No matter how complex is the message generated by the friend, it indicates only his computational abilities and no underlying consciousness. Part of his memory registers are collapsed as the Wigner's learns of the message content, but not the register that stores the outcome of the spin measurement. It makes no sense to consider what the friend has subjectively perceived, and thus there is no paradox of any sort.

The "perception of future" interpretation is intuitively least demanding to comprehend. All the time the friend "really" sees the result and may ponder about it - it is only that the result value does not pertain to the spin alignment at $t_1$, but at $t_3$.

\section{A few words on Extended Wigner's friend experiment}

For the sake of completeness we shall briefly consider the "Extended Wigner's friend" experimental setup of \cite{MW}, although, by now, it should be clear what are our positions also in this particular case. For the details of the setup and the corresponding notation we refer to the article \cite{MW}.

Again, we will start by replacing "Wigner friends" F1 and F2 with some simple quantum automata that undoubtedly undergo linear and unitary evolution. The measurement outcomes of A and W are then beyond any dispute, and no conflicting predictions can occur.

As before, paradoxical interpretations arise only when we consider subjective experiences of the involved friends F1 and F2. And again, we first consider the truly final state of the experiment, after the complete decoherence has occurred and all agents involved are within the same informationally connected reference system. This final stage is necessary not only in any realistic experimental setup, but also in principle, if the agents are to compare the results and discuss their conclusions. Yet again, this final stage, let it be at a moment $t_6$, is ignored both in \cite{MW} and in the subsequent replies, either by omission or by some, surprising to us, tacit agrement on its irrelevance. After this final decoherence, friends F1 and F2 end in well defined brain states, either $|head\rangle$ or $|tail\rangle$, and either $|up\rangle$ or $|down\rangle$, respectively, with all memories and subjective impressions of one that has simply just measured these outcomes at respective times $t_1$ and $t_2$. It is unavoidable that the friends will be confused by the results of interference measurements of A and W, if they look retrospectively and solely from their own perspectives. Neither of the quantum mechanical interpretations that respect universal unitarity can change this fact, and neither finds anything deeply disturbing there. Alternatively, both friends can correctly infer outcome probabilities of interference experiments by assuming Wigner's and Assistant's perspectives, and by accounting for all the relevant information.

It is only insistence on existence of the subjective perceptions of friends before moment $t_6$ that can lead to inconsistencies and to coexistence of conflicting outcomes (like the coexistence of sharp values of otherwise non-commuting observables). While again, there is no objective indication that the corresponding subjective perceptions of the measurement events at $t_1$ and $t_2$ ever existed prior to $t_6$, from the perspective of Wigner, his Assistant and the final decohered system. As before, to avoid the inconsistencies we can freely adopt one of the views: i) these subjective perceptions were "decided" and came into existence at $t_6$ from the W and A point of view, ii) these subjective perceptions (somehow) simply do not exist from the W and A viewpoints, and iii) these subjective perceptions concurrently and continuously existed even from the W and A perspectives, but these were perceptions of the outcomes that would become defined at $t_6$.

A subtle difference between the extended Wigner's experiment and the original one is that the subsequent measurements of Wigner and his Assistant this time do alter the quantum states of the friends' isolated subsystems. However, apart from obscuring problems related to unmotivated flip of the friend's subjective perception, we do not see any larger impact of this difference to our conclusions.

\section{Comparison with other views}

We have offered three interpretations - ones we dubbed "consciousness causes collapse", "no objective consciousness" and "perception of future" - which all have in common negating the very existence, from the Wigner's point of view, of the friend's subjective perception of the $t_1$ measurement outcome in between of $t_1$ and $t_2$. All three can be seen as "sub-interpretations" of the Copenhagen interpretation with subjective collapse where the subjectivity of the collapse is applied consistently and to the full extent. To our knowledge these views do not coincide with any of the commonly known interpretations of quantum mechanics. On the other hand, in the terminology of \cite{OC2}, they all correspond to the same mathematical formalism that is also shared by many-worlds, De-Broigle Bohm, QBism, Rovelli's and Copenhagen's interpretation (as well as by any other unitarity-respecting interpretation). Namely, predictions of all these interpretations match in all objectively measurable aspects: they all predict both that Wigner's interference measurement would yield W+ and that Wigner's friend later recount of events would describe simply seeing a single outcome defined at $t_3$. Above we have intentionally included Copenhagen's interpretation in the list since, in spite of not being conclusively defined, we find it more in the spirit of Bohr's and Heisenberg's ideas to put it here than to interpret it as mechanistic-collapse theory as the authors seem to have implied in [OC2].\footnote{We understand that Brukner in \cite{Brukner} shares this sentiment with us.}

The symmetry of different observers expounded in section \ref{sec:whenandwho} may resemble relativity of information-reference systems of the Rovelli's relational quantum mechanics. For more detailed comparison, we again turn to the analysis of the Wigner's friend experiment elaborated in \cite{Brukner}. We agree with Brukner that the view he presented in detail there has much in common, in the first place, with the Rovelli's interpretation.
In absence (to our knowledge) of sources in literature that scrutinize specifically Wigner's proposed experiment from the perspective of this school of thought, we thus take that its view would follow Brukner's in most of the main aspects.

Both the approach of \cite{Brukner} and the ones we advocate agree about the following tenets: firstly, postulates I and II must be preserved, and, secondly, the definite outcome of the spin measurement and the sharp value of W+ interference outcome cannot coexist in the usual sense. From these premises Brukner derives that "there are no facts of the world", but he does not deny (at least not in an obvious manner) the very existence of concurrent subjective perception of the Wigner's friend in between of $t_1$ and $t_2$ from the Wigner's point. Instead, he never seems to doubt (not even from the Wigner's perspective) that at the (absolute time) moment $t_1$ the friend really perceives a single outcome - it is more that the outcome, i.e.\ the observer's record, should be meaningless "per se" and attains significance only relative to the observer, that is, to the corresponding environment. As clarified in \cite{Brukner2}, it is the problem of "jointly assigning truth values to the propositions {\it about} observed outcomes" (italic added by us), not denying the very existence of the observed outcomes. This is also how we understand, for example, the statement \cite{Brukner}: "If we respect that there are no preferred observers, then there is no reason to assume that the 'facts' of one of them are more fundamental than those of the other. But then, the observers' records cannot be comprised of 'facts of the world', independent of the "environment" in which they have occurred." - namely, that the 'facts' are there, that the records have indeed "occurred", but they are meaningless when taken independently of the "environment". It further seems that the author (\cite{Brukner} page 20) accepts - due to the evidence in the form of message - that the "observer had the perfect knowledge about A1" (i.e.\ about the spin outcome) prior to the interference measurement (which means also from the Wigner's viewpoint, as he is the one that receives the message!), and implies, in parenthesis, that the observer forgets this outcome afterwards.\footnote{The impression is additionally reinforced by the comparison made on page 23 with a double slit experiment in which, according to the relational interpretation, electron should be supposed to "know" which path it took {\it during} the interference measurement.} If this is so, then we encounter the problem, already mentioned, of the final observer's memory of the outcome "gained" at $t_3$. This one might (in one half of the cases) turn out to be different from the one he has just forgot. Yet, what then induces this occasional flip, if the interference measurement itself leaves the state of the observer (i.e.\ the friend) intact? There is another subtle unusual aspect of allowing the final memory determined at $t_3$ to be different from the one occurring at $t_1$: the friend "lives through" experience corresponding to the memory obtained at $t_1$, while the "final" memory he only experiences as the past one and has never "felt it" as unfolding in the present time. Or, maybe he subjectively "lived through" both, which is again unusual? Of course, both of these problematic aspects are not of "objective nature", in the sense that the friend after $t_3$ (when he is finally able to give his account of the events) can no longer tell the difference - neither he knows if the flip has occurred or not, nor he can tell if he ever felt this final outcome as "unfolding" in real time (looking retrospectively, memory in both cases feels genuine and the same). Nevertheless, note that both of these philosophical problems arise relative to a single observer (or environment), and thus cannot be avoided by referencing to relativity of "facts" of different observers. Unfortunately, Brukner himself did not in \cite{Brukner, Brukner2} consider the moment $t_3$ of the final decoherence so that our above interpretation of the text might be prone to misunderstanding.

To the contrary of Brukner, as already stated, we do not maintain that our proposed approach has a problem to establish "facts of the world". It is only that we assume properties of the phenomenon of subjective experience (related in particular to its temporal relativity) different from those commonly attributed to it (neither former nor the latter are so far independently proved and probably even cannot be). Depending upon the chosen sub-interpretation, Wigner can even answer, in a consistent way, the question of the current subjective experience of his friend: either this experience i) emerges only at $t_3$, or ii) never, or iii) it concurrently unfolds, but that of the outcome that will be decided only at $t_3$. In any of the cases, there is no coexistence of conflicting perceptions, and thus nothing that precludes us to attribute general - i.e.\ objective - truth values to the corresponding propositions. We think that this is a very clear distinction of our interpretations from the Brukner and Rovelli's views.

In turn, we find that our proposals contradict the statement of the no-go theorem of the paper \cite{Brukner2}, as they satisfy all four of the theorem propositions. The logical explanation we see in the fact that the proof of the theorem implicitly used certain assumptions about the subjective experience - in particular that Wigner's friend perception of $t_1$ outcome certainly exist in between $t_1$ and $t_2$ from Wigner's viewpoint (taking the message the Wigner receives as a sufficient proof of that). The validity of the no-go theorem can be restored if it is extended with a proposition similar to the following: "5. 'Strong emergent-consciousness hypothesis'. Every state of a functioning brain, even when in superposition, gives rise to the corresponding subjective experience".

Of the very concrete differences in the example of the Wigner's friend, we stress that \cite{Brukner} i) presupposes existence of something (friend's definite $t_1$ outcome perception) that never leaves a trace in the Universe and thus cannot be objectively confirmed that had existed and ii) this something then undergoes a sudden flip at an unclear moment and without a convincing cause. Conversely, a in all three of our views, there is never, from Wigner's perspective, subjective perception of a well defined spin projection pertaining to time between $t_1$ and $t_2$, nor is there ever any flip in the friend's memory/experience.

Of all our sub-interpretations, we acknowledge that there is indeed significant similarity of the "no objective consciousness" sub-interpretation with the QBism views on the Wigner's friend experiment. The latter is expounded with great clarity by Fuchs and Stacey in \cite{QBism}. Yet, while the QBism seems to take an agnostic view in this regard and to insist that quantum mechanics should not attempt to "pierce the other agent’s personal experience", we see the "no objective consciousness" concept as a specific trait of the subjective experience {\it per se}, not as the mere inability of quantum mechanics to account for it. Besides, apart from this sort of "agnosticism", in \cite{QBism} we do not find anything that would unambiguously indicate doubting that the very subjective perception of friend about the outcome exists from the Wigner's perspective. Unfortunately, the authors of \cite{QBism} also do not consider the final decoherence at $t_3$, nor they discuss the fact that interference measurement need not alter the state of the isolated system at all - for these would certainly further clarify the parallels and the differences of the interpretations.

As the presented views correspond to Copenhagen interpretation {\it with subjective collapse}, it makes sense also to compare them with the no-collapse interpretations, in particular with the Everet's relative state viewpoint. Namely, if the collapse is only subjective, does it mean that "objectively" the wave function never collapses, and that thus all terms in a superposition, no matter that they did not actualize from our viewpoint, must be kept and that we must attribute ontological existence to them? Indeed, when we analysed the Wigner's gedanken experiment from the viewpoint of the friend, we ascertained that he must assume Wigner's perspective to correctly predict the outcome of the interference measurement. If he, in between of $t_1$ and $t_2$ in his subjective time reference, upon seeing the result, supposes that the wave function has objectively collapsed, he will get wrong predictions for the W+/W- measurement. Does this imply that there is never any state reduction, for there can always be some external observer?

In our opinion, this is not a necessity. Namely, keeping the "unobserved" terms is relevant for making correct predictions only as long as {\it there are} external observers or external experimental setups that make these terms relevant. In the above example, if there was no Wigner or Wigner-device outside of the box, capable and prepared to measure the W+/W- interference, the collapse observed by the friend might have been not only subjective but also objective. The friend might not know whether there is or there is not such a "Wigner" outside, but the Universe "knows". This becomes more natural if one adopts a view that there exists a reality, defined only to certain extent (insomuch to be in agreement with all the "collapsed" classical information), of which our subjective experiences are only particular perspectives, while the laws of quantum mechanics merely enable us to calculate probabilities of what the reality in the next moment might look like. These subjective experiences then take the role of the "facts of the world", while the wave function in such a view does not have ontic existence - it is just a mathematical tool that helps calculate future predictions based on previous outcomes. Thus as soon as the need to keep certain terms of this mathematical tool definitely ceases, we may very well discard them and take that as an "objective collapse". Or, more correctly stated, there was never anything to discard, as the wave-function never truly existed  in the first place - it is just that nobody any more needs these particular terms in his calculation and never will. We emphasise that such a collapse is then objective, but not in the sense of a "mechanical-collapse" which is induced by nonlinear/nonunitary evolution (and thus why we made the distinction in the first place).

This line of reasoning could be elaborated further, in a natural way. As the evolution between the moments of reality (which correspond to measurements in a broad sense) is unitary and deterministic, it introduces no new information into reality. In this sense, to calculate unitary evolution is a calculation of "what happens when nothing new happens". New information emerges at the point of collapse - i.e.\ of the actualization of reality, with higher probability for those outcome-realities with lesser amount of new information.\footnote{Logicall for this view would be also to seek understanding of elementary particles not as something {\it well described} by the unitary irreducible representations of some total symmetry (super)group, but in which the particles {\it are} these representations. For, irreducible representations can be seen as the tiniest, unbreakable, "irreducible" pieces of information that can exist in a Universe with given symmetries.} This appearance of new information, as opposed to overall constancy of the total amount of information in no-collapse interpretations, is a hallmark of (non deterministic) objective collapse models. The same feature we find e.g.\ in QBist views, as elucidated by Fusch and Stacey in \cite{QBism}: "With the action of the agent upon the system, the no-go theorems of Bell and Kochen–Specker assert that something new comes into the world that wasn’t there previously: It is the “outcome,” the unpredictable consequence for the very agent who took the action." Maybe the most notable proponent of similar views was John Archibald Wheeler, to whom is attributed the following statement which well summarizes the idea: “each elementary quantum phenomenon is an elementary act of ‘fact creation’\ ". Indeed, our views here resonate also well with his stance that "the observations of all the participators, past, present and future, join together to define what we call ‘reality’" \cite{Wheeler} and with some other tenets of his "participatory universe" interpretation. However, we stress that this or any other metaphysical extension is by no means necessary for the remainder of the paper.

\section{Conclusion}

The analysis presented here we started by showing that the series of papers \cite{MW, OC1, OC2, DBB}, somewhat in between the lines but still without any doubt, effectively disqualifies the standard, i.e.\ Copenhagen interpretation as a valid approach to QM. Endeavor to find the real roots of these arguments inevitably led us to the phenomenon of subjective experience.

This should not come to much of a surprise, since the inspiration for Wigner to devise his though experiment, which lies in heart of the listed papers, in the first place was exactly to point to the consciousness as the final frontier where the collapse postulate should be invoked. (However, this led him into direction of mechanical-objective collapse {\it caused by the consciousness}, with all hurdles such an idea brings along.) Many other explanations in the literature also revolve around or close to the notion of subjective experience, but mostly fail, or intentionally avoid, to clearly put a finger on it.\footnote{We can not help getting impression that sometimes authors, in an attempt to stay "rational", irrationally try to avoid the word "consciousness" or anything that sounds alike. However, we must not forget that "being rational" entails looking for the simplest and the most explanatory answer, wherever that takes us, and not necessarily for the one that "sounds rational" by the present standards.}

The other key observation we made was about importance of the, generally overlooked, moment of the final decoherence $t_3$ - which gave us insight into the only objective indication of the Wigner friend's subjective experience in this setup. Recognizing that, according to any of the unitarity respecting interpretations of QM, he will certainly have at least one subjective recollection of the spin measurement (and the one that anyhow seems incompatible with the interference measurement results), we naturally casted a doubt about the very existence of the other, earlier subjective experience of the same measurement, the one which is in addition conflicting and unsubstantiated.

It turned out that the conclusions coming from such line of reasoning are absolutely compatible with the standard QM postulates, landing a welcome support for the challenged Copenhagen interpretation (by this, in our eyes, we justified the title of this paper). Moreover, such approach enables us to keep notion of "facts of the world" in conjunction with the standard postulates of quantum mechanics - a result expected to be impossible by some of the authors \cite{Brukner2}. On the other hand, the price of these new views is a necessity to reexamine our assumptions about the nature of the subjective experience itself.

In particular, we have proposed three novel possible answers to the problem of the friend's subjective perception of the spin outcome (prior to the interference measurement), all of which can be seen as "sub-interpretations" of the Copenhagen interpretation with subjective collapse. We dubbed these sub-interpretations: "collapse causes consciousness", "no objective consciousness" and "perception of future". Each of them implies some different properties of the subjective experience phenomenon. Obviously thus, the new interpretations rely on these corresponding hypotheses about consciousness.

But, the less obvious and generally overlooked fact is that also all of the other interpretations rely on their, tacitly accepted assumptions about consciousness.\footnote{Furthermore, based on these unspoken assumptions, certain authors, e.g.\ in  \cite{MW, Deutsch} try to invalidate some of established quantum mechanical interpretations, CI among the rest.} In this sense, we did not introduce any more of unproved assumptions about the elusive phenomenon of subjective experience than any other interpretation of quantum mechanics, it is only that premises that some of the standard interpretations are based upon are taken for granted and thus pass without notice. As we already pointed out, in the context of Wigner's friend (but also in a more general context) it is common to presuppose that subjective experience is a mere byproduct of the calculations performed in the brain, and furthermore, that this even holds for a brain in a quantum-mechanical superposition (what we named the "strong emergent-consciousness hypothesis").

In truth, not only that these implied assumptions are not proven, but, due to the subjective character of subjective experience, it is quite possible that such assertions are in principle impossible to objectively prove\footnote{For the perspective on related open problems in cognitive science we suggest getting informed with the discussions related to Chinese room argument, Mary's room, sufficiency of Turing's test for inferring existence of consciousness, and alike. As the physicists generally seem to be quite unaware of these problems, so the cognitive science researches seem to be mostly equally unaware of the quantum mechanics and its possible relation to their field - at least judged by a very few hits an Internet search for combination "Chinese room" and "quantum mechanics" yields.}. In the terminology of quantum physics, the confronting views on the nature and causes of subjective experience that appear in cognitive science are therefore more like different {\it interpretations} of consciousness, in the sense in which different ontological views on quantum mechanics with the same experimental predictions we call different interpretations.

In this view, our goal is to show that, instead of invalidating certain interpretations of quantum mechanics, Wigner's friend thought experiment points to a link between interpretations of quantum mechanics and of cognitive science, in the sense that some interpretations of quantum mechanics are compatible only with certain interpretations and assumptions related to the nature of subjective experience. For example, the many world interpretation relies on the above "strong emergent-consciousness hypothesis", DBB interpretation requires basic emergent-consciousness hypothesis but with a caveat that the wave-function of the brain is itself insufficient to give raise to subjective perception, Rovelli and Brukner's interpretation seem to require emergent hypothesis but in a way related to "no facts of the world" dictum - which needs further clarification, QBism is mostly agnostic according to \cite{QBism}, whereas each of our three proposed interpretations goes hand in hand with the corresponding specific assumptions about consciousness that we already discussed at length. In addition, all of the interpretations that are related to subjective collapse view (Rovelli's, Brukner's, QBism and these presented here) are dependent or are at least deeply connected with the "other mind" conundrum - the impossibility in principle to objectively prove existence of subjective experience of any agent/thing.
All these connections, best manifested through the Wigner's friend experiment, implicate that the fundamental problems of quantum mechanics and fundamental question about phenomenon of subjective experience seem to be deeply related and inseparable, so that proper approach would be to treat them jointly.

As a side note of this analysis, we comment that (contrary to some expressed opinions, e.g.\ in \cite{OC1}) actual realization of the Wigner's friend thought experiment is not a prerequisite to differentiate between any of the quantum mechanical interpretations. Namely, to summarize previously drawn conclusions: predictions of mechanical-collapse models, if right, would have to show up in much much simpler experiments; all other interpretations share the same formalism and thus agree on Wigner's thought experiment outcomes, which includes also the variants with communication between Wigner and his friend\footnote{As we have discussed when considering effects of the messages sent by the friend, experiments that include signalling could bear nontrivial implications on understanding of the brain functioning and its relation to probabilities of agent decisions. However, such understanding is rather a prerequisite to carry out the Wigner's experiment, and not the knowledge that we could gain from it.}. In addition, possibility that the friend would subjectively experience any sort of blurred or superposed reality can be also dismissed, even experimentally if need be, far more easily than by carrying out the entire Wigner's experiment. Thus, performing the experiment in reality likely would not provide us with any new insight into quantum mechanics. This is to much of our relief, as it would have been very disappointing that such deep and important understandings hinge upon a feat we are so unlikely to ever perform.

In the concluding remark, we would like to give a sort of disclaimer that should accompany all of the papers on the Wigner's friend subject and alike. Namely, all of the reasoning in such papers, including in this one, is based on the premise that experiments of this type are in principle possible. It is not inconceivable that there could exist some essential limitations which would preclude this type of experiments altogether: either regarding the required ideal informational isolation, or regarding necessary connection of the brain-state reading with the subjective experience, or even regarding the specific combination of the two. If so, that would probably void all of the conclusions derived in these analyses. However, in that case, discovering these limitations would be of a great value for understanding of quantum mechanics, subjective experience or both. Ergo, over a half a century after Wigner devised his thought experiment, it remains a great and inspiring theoretical tool for probing not only the quantum mysteries, but also those related to the subjective perception phenomenon.

\end{document}